\documentclass[12pt]{article}
\usepackage{graphics}

\newcommand{\BH}	{{\rm BH}}
\newcommand{\planck}	{{\rm pl}}
\newcommand{\weak}	{{\rm W}}
\newcommand{\DW}	{{\rm DW}}
\newcommand{\TeV}	{{\rm \;TeV}}
\newcommand{\GeV}	{{\rm \;GeV}}
\newcommand{\MeV}	{{\rm \;MeV}}
\newcommand{\keV}	{{\rm \;keV}}
\newcommand{\kg}	{{\rm \;kg}}

\newcommand{\Max}	{{\rm Max}}
\newcommand{\Min}	{{\rm Min}}
\newcommand{\univ}	{{\rm univ}}

\newcommand{\total}	{{\rm total}}
\newcommand{\sph}	{{\rm sph}}
\newcommand{\gnear}
  {{\;\mbox{\raisebox{-1.6mm}{$\sim$}\hspace{-3.2mm}\raisebox{0.7mm}{$>$}}\;}}
\newcommand{\lnear}
  {{\;\mbox{\raisebox{-1.6mm}{$\sim$}\hspace{-3.2mm}\raisebox{0.7mm}{$<$}}\;}}

\newcommand{\fig}[1]	{Figure \ref{#1}}

\newcommand{\PRD}[3]	{{Phys. Rev. D}       {\bf #1}, #2 (#3)}
\newcommand{\PLB}[3]	{{Phys. Lett. B}      {\bf #1}, #2 (#3)}
\newcommand{\PL}[3]	{{Phys. Lett.}        {\bf #1}, #2 (#3)}
\newcommand{\NuP}[3]	{{Nucl. Phys.}        {\bf #1}, #2 (#3)}

\newcommand{\Nature}[3]	{{Phys. Rev. Lett.}   {\bf #1}, #2 (#3)}

\begin{document}
\baselineskip	= 7mm

\begin{flushright}
\begin{minipage}[b]{33mm} 
 DPNU-98-19\\
 hep-ph/9805455\\
 May 1998
\end{minipage}
\end{flushright}

\begin{center}
{\Large\bf Black Hole Baryogenesis}

\vspace{10mm}

{\large Yukinori Nagatani}
\footnote{E-mail: nagatani@eken.phys.nagoya-u.ac.jp}\\
{\it Department of Physics, Nagoya University, Nagoya 464-8602, Japan}

\end{center}

\vspace{10mm}

\begin{center} {\bf ABSTRACT} \end{center}
\begin{quotation}
We find that the Hawking radiation of black holes
can create the baryon number
in the Higgs phase vacuum of the extended standard model
with 2-Higgs doublets.
We propose new scenarios of baryogenesis based on the black holes,
and one of our scenarios
can explain the origin of baryon number in our universe,
if most of the matter existed as primordial black holes.
\end{quotation}

\newpage


The thermal radiation from a black hole was discovered
by Hawking \cite{Hawking},
and contributions of primordial black holes in the early universe
have been discussed \cite{Primordial}.
Then it is natural to ask if black holes can play a role in the baryogenesis.
There have been proposed
many scenarios of the baryogenesis:
Among those, the electroweak baryogenesis proposed by Cohen et al. \cite{CKN}
is an important scenario,
in which the electroweak domain wall
created by the first order phase transition
plays a crucial role.

In this paper,
we find formation of the electroweak domain wall surrounding the black hole,
and we show that the Hawking radiation from the black holes
can create the baryon number
in the Higgs phase vacuum of the extended standard model (SM)
with two Higgs doublets.
We then propose three scenarios of the baryogenesis by the black holes:
one of our scenarios can create the baryon number
with the baryon-entropy ratio up to $B/S \simeq 10^{-9}$
in the early universe,
and can satisfy the requirement of the big-bang nucleosynthesis (BBN),
if most of the matter existed as primordial black holes
with mass of some hundreds kilograms.
The most important difference
between our scenario and the ordinary one \cite{CKN}
is the requirement on the nature of the phase transition:
In the ordinary scenario, one needs the first order phase transition to
create the domain wall to realize non-equilibrium,
and the structure of the domain wall is determined by the
dynamics of phase transition,
while in our scenario, {\it we do not need the first order phase transition},
since the thermal structure of the black hole creates the domain wall
and determines its structure.


First,
we discuss how to satisfy the Sakharov's three conditions \cite{Sakharov}
for the baryogenesis.
Let us consider the Schwarzschild black holes
whose Hawking temperature $T_\BH$ is higher than
the critical temperature of electroweak phase transition.
In the Higgs phase vacuum,
the radiation from these black holes restores the electroweak symmetry
at the neighborhood of the horizon
and the electroweak domain wall does appear
as we shall demonstrate shortly.
Then we can discuss the baryogenesis scenarios
in analogy with the ordinary electroweak baryogenesis \cite{CKN}.
Here
we assume the two-Higgs-doublets extension of the standard model (2HSM)
as the background field theory for the origin of CP phase in the domain walls,
and that the electroweak phase transition is the second order
with its critical temperature taken as $T_\weak = 100 \GeV$
for simplicity.
Actually,
the Sakharov's  three conditions \cite{Sakharov} for baryogenesis
are satisfied as follows:
\begin{enumerate}
 \item The baryon number violation: it can occur
       either through (a) or (b) processes:
       \begin{enumerate}
	\item The sphaleron process \cite{Sph} takes place
	      in the symmetric region
	      or in the domain wall near the symmetric region.
	\item The interaction with the black hole
	      (exchange, falling into and radiation of particles)
	      is the baryon-number-violating process.
       \end{enumerate}
 \item The C-asymmetry: the SM is the chiral theory.\\
       The CP-asymmetry: Here we assume in 2HSM that
       the domain wall has the space-dependent CP phase.
 \item Out of equilibrium:
       The black-hole radiation is a non-equilibrium process.
       This radiation creates the spherical domain wall
       and the radiated particles pass through.
\end{enumerate}


To begin with we note that
the Schwarzschild black hole mass $m_\BH$, temperature $T_\BH$,
Schwarzschild radius $r_\BH$, and lifetime $\tau_\BH$
are related by the equations:
$
 T_\BH		=  \frac{1}{8 \pi} \frac{m_\planck^2}{m_\BH},\ 
 r_\BH		= 2 \frac{m_\BH}{m_\planck^2}
 		= \frac{1}{4 \pi} \frac{1}{T_\BH},\ 
 \tau_\BH	\simeq  \frac{10240}{g_*} \frac{m_\BH^3}{m_\planck^4}
		= \frac{20}{\pi^2 g_*} \frac{m_\planck^2}{T_\BH^3},
$
where $g_*$ is the freedom of the massless particles that
this black hole can decay into at its temperature.
In the electroweak critical temperature, 
we have $g_* \simeq 100$.
In this paper,
we parameterize black holes by its Hawking temperature
rather than its mass for convenience.
We display these relations in \fig{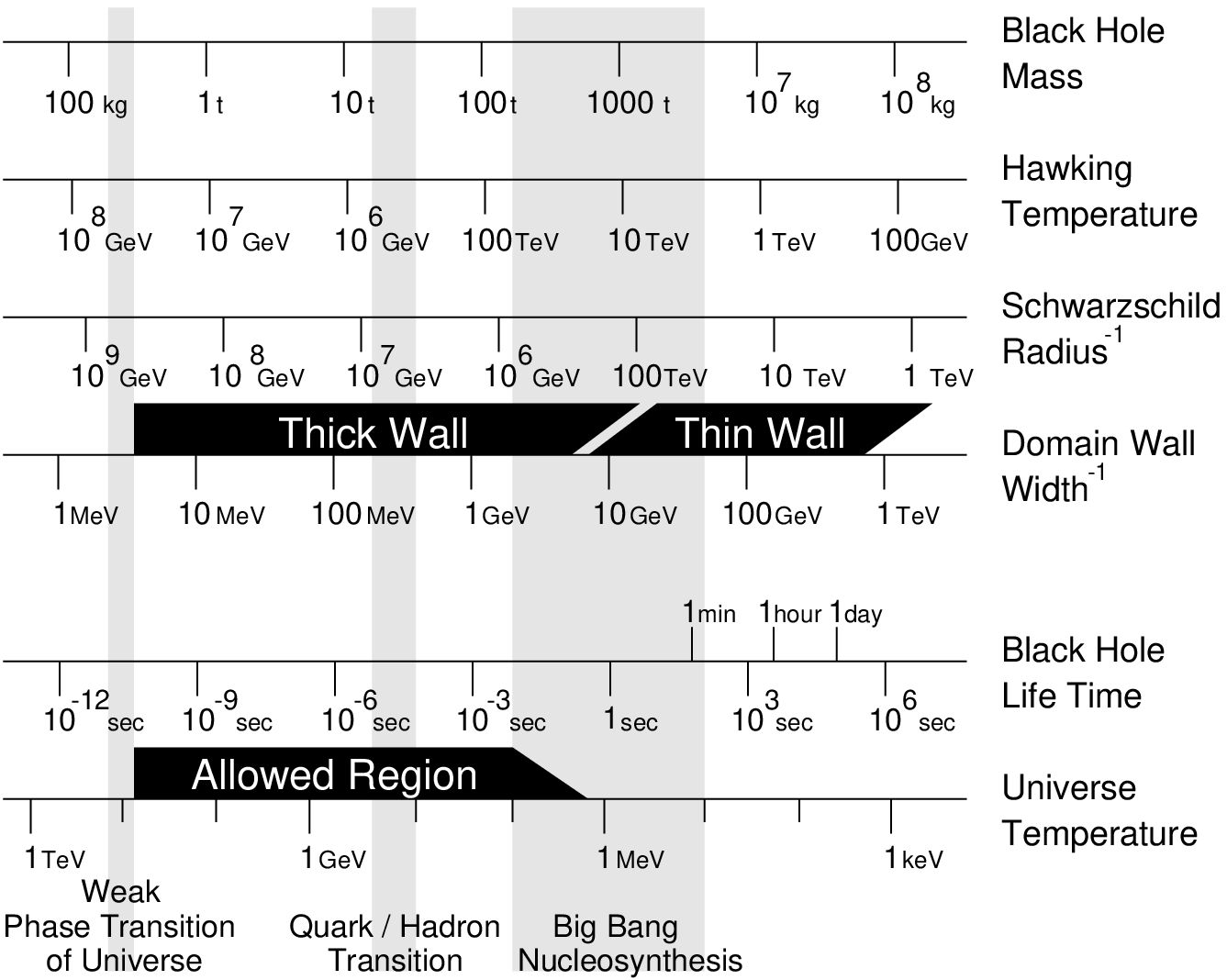}.

\begin{figure}[htbp]
\begin{center}%
 \ \includegraphics{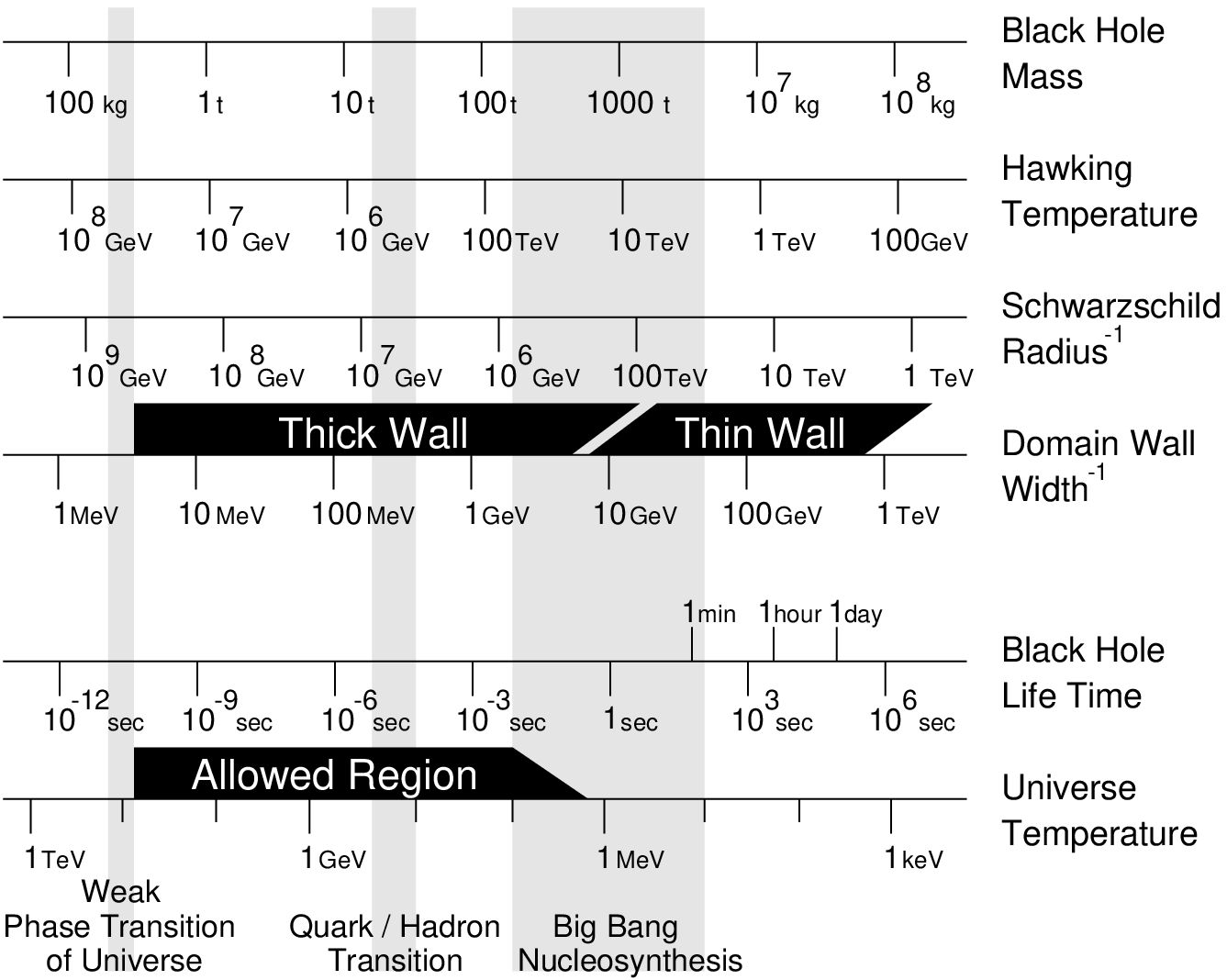}
 \caption{The relations between parameters of the Schwarzschild black hole.
 	  In the relation between the black-hole lifetime and
 	  the universe temperature,
 	  we used correspondence between
 	  the black-hole lifetime and the age of the universe,
 	  because we assumed that
 	  many primordial black holes had existed in the very early universe
 	  and evaporated at the age of the universe.
          }%
          \label{BHTmpLet.eps}%
\end{center}%
\end{figure}

Now,
we consider the space-dependence of the temperature
in the neighborhood of the black hole
as a quasi-stationary thermal equilibrium picture.
Because of the spherical symmetry of the black hole,
we put the local temperature as $T(r)$,
where $r$ means the radius from the black hole center.
We assume that a sphere with the radius $r \ (r \gg r_\BH)$
has the Planck radiation at the temperature $T(r)$
because of the heating-up by the Hawking radiation.
The black hole has the Hawking radiation with the total energy flux being
$ J_\BH = \frac{\pi^2}{120} g_* T_\BH^4 \times \pi r_\BH^2 $
per unit time,
but the total energy flux on the sphere with radius $r$ is
$ J(r) = \frac{\pi^2}{120} g_* T(r)^{4} \times \pi r^2. $
Then the conservation of the energy flux tells us that
$
 T(r) \simeq T_\BH \left(\frac{r_\BH}{r}\right) ^{1/2}, 
$
where we approximated $g_*(T(r)) \simeq g_*(T_\BH)$.

Near the black hole horizon $(r \sim O(r_\BH))$,
we must also take account of the red-shift effect \cite{Landau}.
Then we assume the local temperature as
$
 T(r) \simeq \frac{1}{\sqrt{g_{00}(r)}} T_\BH
	\left(\frac{r_\BH}{r}\right)^{1/2}
      = T_\BH \left(\frac{r_\BH}{d}\right)^{1/2}, 
$
where $g_{00}(r) = 1 - r_\BH / r$ is an element of Schwarzschild metric,
and we defined the distance $d$ from the black hole horizon as $r = r_\BH + d$
in the last equality.
Since our baryogenesis scenarios depend only on the low temperature region
like the electroweak scale,
we do not need to consider this red-shift effect,
when the temperature of black hole is greater
than the electroweak temperature.
In the background with finite temperature $T_{\rm bg}$
$(T_{\rm bg} \ll T_\weak \ll T_\BH )$ as in the early universe,
these analyses result in
\begin{eqnarray}
 T(r_\BH + d) &\simeq&
  T_\BH \left[\;
	 \left( \frac{r_\BH}{d} \right)^2 +
	 \left( \frac{T_{\rm bg}}{T_\BH} \right)^4
	 \; \right] ^{1/4}. \label{eqTemp3}
\end{eqnarray}

Let us discuss formation of the domain wall.
The local temperature $T(r)$
is a decreasing function of $r$.
This local temperature configuration allows us
a nontrivial phase structure of the symmetry breaking depending on the space,
i.e., the vacuum expectation value (VEV) of Higgs doublets
depends on the distance from the center of the black hole $r$.
By the local thermal equilibrium
the space dependence of the Higgs doublets may take the form
\begin{eqnarray}
 \langle\phi_i(r)\rangle &=& \langle\phi_i\rangle_{T=T(r)}.
\end{eqnarray}
For simplicity,
we assume that the electroweak phase transition is the second order
and the simplest form of the Higgs VEV as
$
 (|\langle\phi\rangle_T|/v)^2 + (T/T_\weak)^2 = 1.
$
Then the CP-phased neutral Higgs VEV
may be written as
\begin{eqnarray}
 \langle\phi_1^0(r)\rangle &=&
  \left\{
   \begin{array}{lcl}
    0 & & (r \leq r_\DW) \\
    v_1 f(r) \; e^{-i \Delta\theta (1-f(r))} & & (r > r_\DW)
   \end{array}
  \right.,
\end{eqnarray}
where $f(r) = \sqrt{1 - (T(r)/T_\weak)^2}$ (see \fig{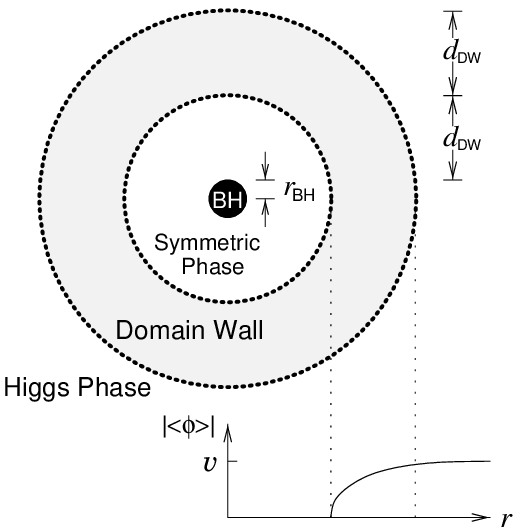}).
In this configuration of the Higgs VEV,
the width of our domain wall $d_\DW$
is equal to the depth of the symmetric region.
By $T(r_\BH + d_\DW) = T_{\weak}$, we find
\begin{eqnarray}
 d_\DW
  &\simeq& r_\BH \left( \frac{T_\BH}{T_\weak} \right)^2
  \;=\; \frac{1}{4 \pi} \frac{T_\BH}{T_\weak^2}.
\end{eqnarray}
We can understand that this electroweak phase structure is determined by
the thermal structure of the black hole.
We illustrate this width
together with other black-hole parameters in \fig{BHTmpLet.eps}.

\begin{figure}[htbp]
\begin{center}%
 \ \includegraphics{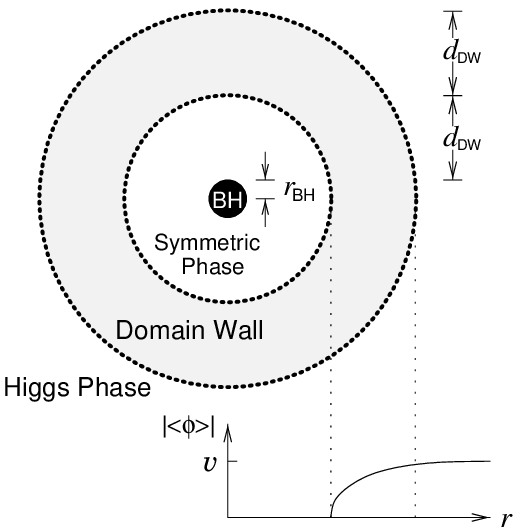}
 \caption{The electroweak phase structure depending on the space
	  near the black hole (In this figure $T_\BH \simeq 0.3 \TeV$).
          }%
          \label{WBH.eps}%
\end{center}%
\end{figure}

On the other hand,
in the ordinary electroweak baryogenesis scenarios \cite{CKN},
structure of the domain wall
is determined by the dynamics of phase structure,
and they assumed $f(z) = \frac{1}{2}[\tanh{z/\delta} + 1]$,
where $z$ is the perpendicular direction to the domain wall and
$\delta$ is the width of the domain wall.

Our quasi-stationary thermal equilibrium picture is valid
when the black-hole lifetime is large enough
to keep the stationary weak domain wall: $1 \ll \tau_\BH/r_\DW$.
If we use $g_* \simeq 100$ for $T > T_\weak$, 
then we get a restriction for the black-hole temperature as
$T_\BH^\Max \ll 2.5 \times 10^{10} \GeV$.
For every black hole which satisfies this restriction,
the temperature rises gradually in time,
and the depth of the wall increases.
Eventually, this requirement for the black hole gets broken and 
hence the depth of the black-hole domain wall is maximized as
$r_\DW^{\Max} = \tau_\BH^{\Min} \simeq (5.0 \keV)^{-1}$.

The black hole also radiates entropy by the Hawking radiation.
The total entropy flux
on the black-hole horizon is
$
 J_S^{\rm horizon} = -\frac{d}{dt} S_\BH = \frac{\pi}{480} g_* T_\BH,
$
where $S_\BH = A_\BH m_\planck^2 / 4$
is the Beckenstein entropy of the black hole \cite{Beckenstein},
where $A_\BH$ is the area of the black-hole horizon.
But the radiation process is a kind of diffusion,
and hence
the total entropy flux on the sphere with the radius $r$ may be written as
$
 J_S(r) = J_S^{\rm horizon} \frac{T_\BH}{T(r)},
$
because the total entropy flux on the thermal sphere
with the temperature $T(r)$ and with the area $A(r)$ is
$J_s \propto T(r)^3 A(r)$, while the conserved energy flux is
$J \propto T(r)^4 A(r) = $ const.
In the universe with temperature $T_\univ$,
the contribution of the entropy from the black hole to the universe is
\begin{eqnarray}
 J_S^\univ &=& J_S^{\rm horizon} \; \frac{T_\BH}{T_\univ}
  \;=\; \frac{\pi}{480} g_* \; \frac{T_\BH^2}{T_\univ}. \label{EntropyFlux}
\end{eqnarray}


Here
we propose three scenarios of the baryogenesis due to these black holes:
The first one is a kind of spontaneous baryogenesis scenario
in the domain wall \cite{CKN},
which we call the ``thick-wall black-hole baryogenesis''.
The second is a variant of the charge-transport scenario \cite{CKN},
which we call the ``thin-wall black-hole baryogenesis''.
The third is another variant of the charge-transport scenario,
but the hypercharge is transported from the domain wall to the black hole
rather than to the symmetric region.
We call this the ``direct black-hole baryogenesis''.


{\it Thick-wall black-hole baryogenesis.} ---
When width of the domain wall is larger than the mean free path of
the top quark:
\begin{eqnarray}
 d_\DW > l_q \sim (10 \GeV)^{-1} 
 &\leftrightarrow& T_\BH > 4\pi T_\weak^2 l_q \sim 12 \TeV,
\end{eqnarray}
we can consider the spontaneous baryogenesis scenario
as in the ordinary electroweak baryogenesis.
The C- and CP-asymmetries take place in the domain wall
as the space-dependent physical CP phase of the domain wall
and the baryon-number-violating process also occurs in the domain wall near
the symmetric phase as the sphaleron one.
In the domain wall or broken phase, the sphaleron transition rate is
$
 \Gamma_\sph = \kappa \alpha_\weak T^4 e^{-E_\sph/T},\ 
 E_\sph(T) = \frac{2M_W(T)B}{\alpha_\weak} \simeq
  \frac{|\left<\phi\right>|_T}{v} \times 10 \TeV,
$
where $\kappa \sim O(1)$ is a numerical constant.
When $\left<\phi\right>/v < \epsilon \simeq 1/100$,
i.e., $E_\sph < T_\weak$,
the sphaleron transition is not suppressed by the exponential factor.
Then we can consider the sphaleron process
only in the neighborhood of the symmetric region in the domain wall.
We write the width of this region as
$\epsilon^2 d_\DW$ because of the form of Higgs VEV.
Then the volume of the sphaleron transition at work is
$ V = 4\pi d_\DW^2 \times \epsilon^2 d_\DW $.
The space-dependent CP phase, or equivalently the time-dependent one for
the radiated particles, is
$
 \dot{\theta} \simeq \theta' =
  \epsilon \frac{\Delta\theta}{\epsilon^2 \:d_\DW}
  = 4\pi \frac{\Delta\theta}{\epsilon} \frac{T_\weak^2}{T_\BH}.
$
The relation between the baryon number chemical potential
and the time-dependent CP phase is
$ \mu_B = {\cal N} \dot{\theta} $,
where ${\cal N} \sim O(1)$ is a model-dependent constant \cite{CKN}.
Finally, we can write down
by the detailed-balance relation
the rate of the baryon number creation per black hole:
\begin{eqnarray}
 \dot{B} &=& - V \; \frac{\Gamma_\sph}{T_\weak} \; \mu_B
 \;=\; \frac{{\cal N}\epsilon\Delta\theta}{4\pi}
       \kappa \alpha_\weak^4 \frac{T_\BH^2}{T_\weak}.
\end{eqnarray}


{\it Thin-wall black-hole baryogenesis.} ---
When width of the domain wall is smaller than the mean free path of
the top quark:
$d_\DW < l_q \leftrightarrow
T_\BH < 4\pi T_\weak^2 l_q \sim 12 \TeV$,
we can consider the charge-transport scenario
in the ordinary electroweak baryogenesis \cite{CKN}.
The C- and CP-asymmetries take place in the domain wall,
while the baryon-number-violating process does in the symmetric region
inside the black-hole domain wall.
The CP-violated domain wall has
non-zero reflection ratio for the hypercharge.
Then there is non-zero hypercharge flux $F_Y$
from the domain wall to the symmetric region.
We use the sphaleron rate per unit time per unit volume as
$
 \Gamma_\sph = \kappa \alpha_\weak^4 T^4. 
$
The volume of the region where sphaleron process is at work is
$
 V = \frac{4\pi}{3} (r_\BH + d_\DW)^3 - \frac{4\pi}{3} r_\BH^3
   = \frac{1}{16 \pi^2} \frac{1}{T_\weak^3} f\left(T_\BH/T_\weak\right),
$
where we defined
$
 f[X] = X^{-1} + X + \frac{1}{3}X^3.
$
Finally, the baryon number creating rate per black hole is
\begin{eqnarray}
 \dot{B} &=& -V \frac{\Gamma_\sph}{T_\weak} \mu_B
   \simeq - \frac{1}{16\pi^2} \frac{\kappa \: \alpha_\weak^4}{5/3+N_H} \:
   \frac{F_Y}{T_\weak^2} f\left(\frac{T_\BH}{T_\weak}\right),
\end{eqnarray}
with $N_H$ being the number of Higgs doublets,
where we used a relation between hypercharge density and baryon chemical
potential, $Y = (5/3+N_H) T^2 \mu_B$,
and we also used $Y \simeq F_Y$, because the depth of the symmetric region is
smaller than the mean free path of the top quark.


{\it Direct black-hole baryogenesis.} ---
The interactions with the black holes
are fully baryon-number-violating process
because of the ``no-hair theorem''.
The process of the particle exchange with the black hole,
with the particles falling into and radiating from the black hole,
are the processes conserving energy, spin and charges but
are the baryon-number-violating processes.
We take the cross section for the massless particles 
as a shadow area of the black hole
$ \sigma_\BH = \pi r_\BH^2 $.
Then the baryon-number-violating rate in one black hole
for the stable observers at an infinite distance is
$
 \tilde{\Gamma}_\BH \simeq n \sigma_\BH |v| =
 \frac{3}{4} \frac{1}{r_\BH} = 3\pi T_\BH, 
$
where $n = (\frac{4\pi}{3} r_\BH^{\;3})^{-1}$
 is the number density of the black hole
and $v = 1$ is the velocity of massless particles
falling into the black hole.
Since we consider the black hole is a perfectly thermal and black object,
the baryon number creation per unit time for the observer is
$
\dot{B} = - \frac{\tilde{\Gamma}_\BH}{T_\BH} \mu_B, 
$
where $\mu_B$ is chemical potential for baryons
in the neighborhood of the black hole,
$
\mu_B = \frac{Y}{T_\BH^2} \; \frac{1}{5/3+N_H}, 
$
and $Y \simeq F_Y$ means the hypercharge density near the black hole
as the atmosphere of the black hole.
In these relations, we used the black-hole temperature $T_\BH$ rather than
the local temperature $T(r)$,
because this phenomenon may be the property of the black hole itself
for the observer at an infinite distance and may be characterized
by the black-hole temperature.
Finally, the baryon number creating rate by this process is
\begin{eqnarray}
  \dot{B} &=& - \frac{3\pi}{5/3+N_H} \frac{1}{T_\BH^{\;2}} \; F_Y.
\end{eqnarray}


In the early universe,
we assume that most of the matter existed as the primordial black holes
and evaporated through creating baryons in our processes.
This assumption tells us a relation between the lifetime of the black holes
and the age of the universe when the baryon number was created:
$
 t_\univ \simeq \tau_\BH.
$
In our theory,
the evaporation of the black hole in the Higgs phase vacuum
is essential to creating baryons.
Then the lifetime of the black holes must be
greater than the age of the universe
at the electroweak phase transition $T_\univ=T_\weak$ 
in the standard cosmology.
By the relation between the age of the universe and the temperature of
the universe in the radiation dominant era \cite{KolbTurner}:
$
 t_\univ = 0.301 \; \frac{1}{\sqrt{g_*}}
  \frac{m_\planck}{T_\univ^2},
$
this restriction gives us
a lower bound for the black-hole mass $m_\BH \gnear 250\kg$
and an upper bound for the black-hole temperature
$T_\BH \lnear 4.2 \times 10^7 \GeV$
(see \fig{BHTmpLet.eps}).

If we disregard the possibility of the nucleosynthesis
near the black hole due to the black hole,
we must require that
our theory does not affect
the very successful BBN theory \cite{KolbTurner}
at $T_\univ \simeq 10 \MeV \sim 0.1 \MeV$.
Then we have a lower bound
for the black-hole temperature $T_\BH \gnear 2.0\TeV$
(see \fig{BHTmpLet.eps}).


Recalling the total out-going flux of entropy (\ref{EntropyFlux}),
we have the baryon-entropy ratios in the radiations:
\begin{eqnarray}
 \frac{\dot{B}}{J_S^{\total}} &=&
    \displaystyle
    -\frac{120}{\pi g_*} \; \epsilon {\cal N} \Delta\theta \:
     \kappa\alpha_\weak^4
     \frac{T_\univ}{T_\weak}
     \qquad \mbox{(thick wall)}
\end{eqnarray}
for $4.2 \times 10^7 \GeV \gnear T_\BH \gnear 12 \TeV$
as a thick domain-wall effect, while
\begin{eqnarray}
 \frac{\dot{B}}{J_S^{\total}} &=&
  \left\{
   \begin{array}{ll}
    \displaystyle
    -\frac{30}{\pi^3 g_*} \frac{\kappa \: \alpha_\weak^4}{5/3+N_H}
     \frac{T_\univ \; F_Y}{T_\BH^2 T_\weak^2}
     f\left(\frac{T_\BH}{T_\weak}\right)
     & \mbox{(thin wall)} \\[5mm]
    \displaystyle
    -\frac{1440}{g_*} \frac{1}{5/3+N_H}
     \frac{T_\univ \; F_Y}{T_\BH^4}
     & \mbox{(direct)}.
   \end{array}
  \right.
\end{eqnarray}
for $12 \TeV \gnear T_\BH \gnear 100 \GeV$ as a thin domain-wall effect.
These values are not exactly $B/S$ but are equal to the $B/S$
on the order of magnitude.
We plot these values as $\Delta\theta = \pi$ and
top mass $m_{\rm top}=174\GeV$
in \fig{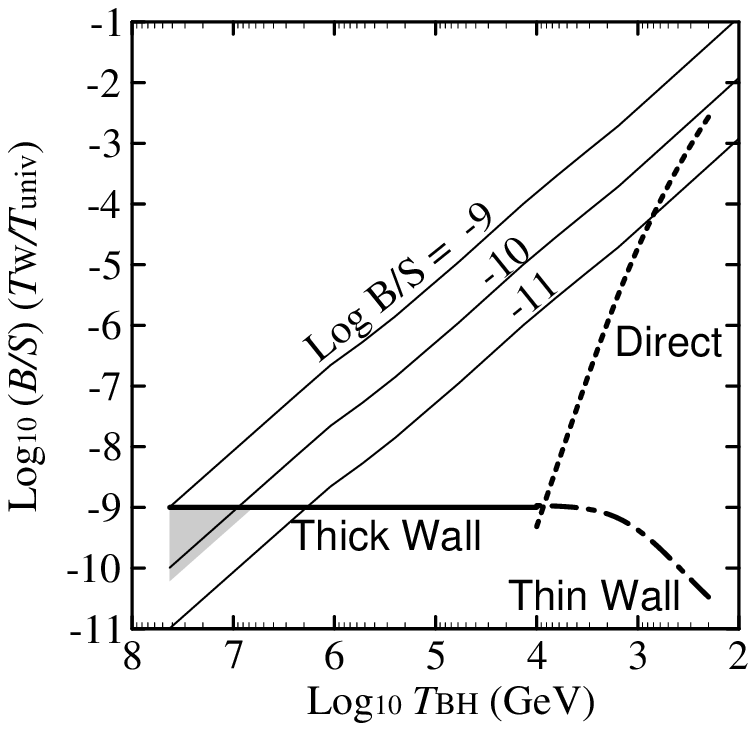}.

\begin{figure}[htbp]
\begin{center}%
 \ \includegraphics{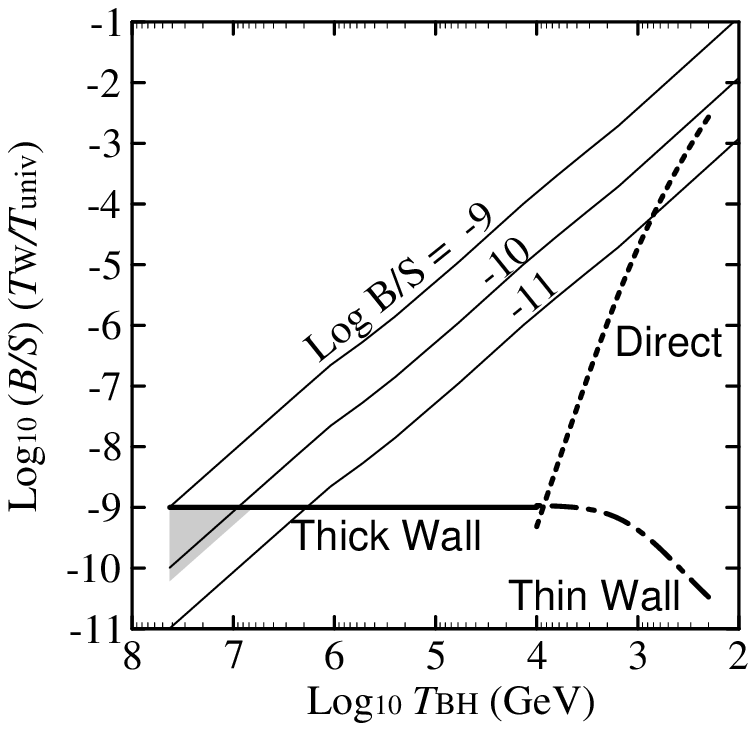}%
 \caption{The resultant baryon-entropy ratio
 	  times universe temperature correction
 	  versus black-hole temperature.
 	  In the curves of $B/S={\rm const}$,
 	  we assumed correspondence between
 	  the black-hole lifetime and age of the universe,
 	  as many black holes evaporated in the universe.
 	  The shaded region satisfies the BBN requirement.
          }%
          \label{ResultLet.eps}%
\end{center}%
\end{figure}

Finally,
the baryon-entropy ratio
in our {\it thick-wall black-hole baryogenesis scenario}
can satisfy the BBN requirement $B/S \sim 10^{-10}$
\cite{KolbTurner}
when $4.2 \times 10^7 \GeV \gnear T_\BH \gnear 1 \times 10^7 \GeV$,
and we can obtain $B/S \sim 10^{-9}$
when $T_\BH \simeq 4.2\times10^7$.
In other words,
if the mean mass of the primordial black holes is some hundreds kilograms,
then we can explain the baryon-entropy ratio by our mechanism.

In conclusion,
we have proposed a new scenario of the baryogenesis
which does not need the first order phase transition,
but does require the primordial black holes.


I would like to thank Prof. K. Yamawaki and Prof. A. I. Sanda
for helpful suggestions and discussions,
and also for careful reading the manuscript.
I also appreciate helpful suggestions
of Prof. A. Nakayama and K. Shigetomi.
The work is supported in part
by a Grant-in-Aid for Scientific Research
from the Ministry of Education, Science and Culture
(No. 80003276).


{\it Note added}:
After submitting this paper, I was informed of the paper \cite{Barrow}
which discussed the baryogenesis based on the Grand Unified Theory (GUT)
by primordial black holes.


\end{document}